\documentstyle[11pt,aaspp4,epsfig]{article}

\def\Msun{\hbox{$M_{\odot}$}}

\lefthead{Cunha et al.}
\righthead{Light Element Abundance Patterns in the Orion Association: I)
HST Observations of Boron in G-dwarfs}

\begin{document}

\title{Light Element Abundance Patterns in the Orion Association: I) HST
Observations of Boron in G-dwarfs}

\author{Katia Cunha}
\affil{Department of Physics, University of Texas at El Paso, El Paso, TX
79968}
\affil{Observat\'orio Nacional - CNPq, Rio de Janeiro, Brazil}
\affil{katia@baade.physics.utep.edu}

\author{Verne V. Smith}
\affil{Department of Physics, University of Texas El Paso, El Paso, TX 79968}
\affil{verne@barium.physics.utep.edu}

\author{Etienne Parizot}
\affil{Dublin Institute for Advanced Studies, Dublin, Ireland}
\affil{parizot@cp.dias.ie}

\author{David L. Lambert}
\affil{Department of Astronomy, University of Texas, Austin, TX 78712-1083}
\affil{dll@anchor.as.utexas.edu}

\begin{abstract}
The boron abundances for two young solar-type members of the Orion
association, BD\,-6$^{\circ}$\,1250 and HD\,294297, are derived from
HST STIS spectra of the B I transition at 2496.771 \AA. The best-fit
boron abundances for the target stars are 0.13 and 0.44 dex lower than
the solar meteoritic value of log $\epsilon$(B)=2.78.  An
anticorrelation of boron and oxygen is found for Orion when these
results are added to previous abundances obtained for 4 B-type stars and the
G-type star BD\,-5$^{\circ}$\,1317.  An analysis of the uncertainties
in the abundance calculations indicates that the observed
anticorrelation is probably real.  The B versus O relation observed in
the Orion association does not follow the positive correlation of
boron versus oxygen which is observed for the field stars with roughly
solar metallicity.  The observed anticorrelation can be accounted for
by a simple model in which two poorly mixed components of gas
(supernova ejecta and boron-enriched ambient medium) contribute to the
new stars that form within the lifetime of the association.  This
model predicts an anticorrelation for Be as well, at least as strong
as for boron.
\end{abstract}

\newpage
\section{Introduction}
\label{introduction}

The first results for boron abundances in stars of the Orion
association were provided by a study of the B\,{\sc ii} 1362 \AA\ line
in four B-type stars (Cunha et al.  1997).  The non-LTE abundances
derived for these Orion stars were between log $\epsilon$(B) = 2.5 -
2.9, i.e. close to the boron abundance measured in solar system
meteorites (log $\epsilon$(B) = 2.78) or the solar photosphere (log
$\epsilon$(B) = 2.7, where log $\epsilon$(X)=log N(X)/N(H) + 12).  
A second study of boron in the Orion
association using the G-dwarf BD-5$^{\circ}$\,1317 found a boron
abundance lower by about 0.7 dex than the solar/meteoritic value
(Cunha, Smith \& Lambert 1999).  This particular Orion member was
picked because i) it had an undepleted lithium abundance and ii) it
exhibited one of the largest oxygen abundances measured in the
association (log $\epsilon$(O)=9.11; Cunha, Smith \& Lambert 1998).
The absence of Li depletion guarantees that the boron was not
destroyed in the stellar interior.  As a consequence, the observed
boron abundance is expected to reflect the original abundance at the
time of star formation.  As for the high oxygen abundance, it is
attributed to the self-enrichment of the Orion cloud (Cunha \& Lambert
1994), caused by the explosion of massive stars as type II supernovae
(SN II).  According to this model, the O-rich SN II ejecta `pollute'
certain pockets of gas around the site of the explosion, from which
new stars form with enhanced oxygen abundances.  This idea is
corroborated by the fact that the most O-rich stars in the Orion
association are also the youngest members and generally co-located
(Cunha et al. 1998).

In the present study, two additional stars are examined.  They consist
of G-dwarfs whose radial velocities and proper motions indicate that
they are definite members of the Orion association (Cunha, Smith \&
Lambert 1995).  These two stars also show different degrees of oxygen
enrichment and fulfill the requirement that they do not show
significant Li depletion.  They can thus be used to study the
self-enrichment process within the Orion clouds.  This is an important
goal from the theoretical point of view, since SN II's are known to be
the source not only of O, but also indirectly of LiBeB. According to
current knowledge (e.g. Vangioni-Flam et al.  2000), beryllium and
boron can be synthesized only by spallative processes in which heavier
nuclei (mainly C, N and O) are spallated in reactions induced by
energetic particles (EPs).  Two distinct mechanisms have been invoked,
depending on whether the EPs are nuclei or neutrinos.  In the first
case, B (and Li and Be) is produced by the spallation of C, N, and O
by protons and $\alpha$'s.  The other mechanism, known as the
$\nu$-process (or neutrino-spallation), consists in breaking nuclei
with energetic neutrinos released in enormous numbers during the
explosion of Type II Supernovae.  The carbon nuclei spalled in this
process to make $^{11}$B are those synthesized by the massive SN II
progenitor.  As a consequence, the $\nu$-process boron is
`ready-mixed' with the CNO in the SN II ejecta and all these nuclei
are released together into the ambient medium.  The $\nu$-process has
been invoked to cope with a classical problem resulting from the
confrontation of the observed (solar system) values of the B/Be and
$^{11}$B/$^{10}$B ratios with those predicted by the nucleo-spallation
process alone that produces too little $^7$Li and $^{11}$B. The
$\nu$-process is predicted to produce $^{7}$Li and $^{11}$B but
virtually no $^{6}$Li, no $^{9}$Be and no $^{10}$B. Therefore, it
offers a way to resolve the classical problem.

From the detailed analyses of Be and B abundances in metal-poor halo
stars, it has been recently proposed that the EPs responsible for most
of the light element production are probably accelerated inside
superbubbles created by repeated SN II's in OB associations (Parizot
\& Drury, 1999, 2000; Parizot, 2000; Ramaty et al.  2000).  Since it
is known that the Orion association blew just such a superbubble (SB)
from the winds and explosions of massive stars in subgroup Ia (e.g.
Brown et al.  1995), it may be expected that significant Li, Be and B
production occurred recently in or close to the Orion molecular cloud. 
The self-enrichment in O of the Orion stellar association should
therefore be accompanied by a significant enrichment in light elements
as well, and the relation between the two processes is worth studying.

In the last decade, B abundances have been measured in disk and halo
stars (Duncan, Lambert \& Lemke 1992; Duncan et al.  1997;
Garc\'{\i}a-L\'{o}pez et al.  1998; Primas et al.  1999 and Cunha et
al.  2000) showing a global positive correlation of boron with oxygen
which is understood as the progressive enrichment of the interstellar
medium (ISM) in both metals and light elements.  However, the
situation in local environments has never been studied, and one may
ask whether the same kind of correlation also holds on smaller scales,
within individual stellar associations.  Our study intends to answer
the question in the case of Orion.

\section{STIS Observations and Spectra}

The two Orion targets, BD\,-6$^{\circ}$\,1250 and HD~294297, were
observed with the STIS spectrograph on the Hubble Space Telescope
(HST), over 10 and 11 orbits each, which were needed in order to
obtain combined spectra with a S/N $\sim$50 for the analysis of B I
at 2496.771 \AA. (The measured S/N across the spectral region synthesized
ranged from 40-60 in both stars).
The observations were obtained with the MAMA detector
in the ACCUM mode with the first order grating G230M, plus 52X0.1 slit
to record the spectral region between 2454 and 2545 \AA\ with a
resolution R$\sim$14,000.

The observed STIS spectra were processed by the pipeline calibration
from STScI with the IRAF package `calstis' which was used in order to
subtract the bias, divide by the flatfield and wavelength calibrate
the data.  These pipeline calibrated spectra were then combined with
the IRAF task 'mscombine' and extracted one-dimensional spectra were
obtained with the task 'apsum'.  Unlike the spectra obtained with the
echelle grating, the levels of scattered light in the first-order
grating spectra are insignificant and the step of subtraction of
scattered light was not needed.  The final spectra of the two Orion
targets are shown in the two panels of Figure 1.

\section{Analysis}

The stellar parameters adopted in the calculation of the model
atmospheres for the two studied stars (plus a sample of Orion members)
are presented in Table 1.  The  effective temperatures  from
Cunha et al. (1995)
were obtained from a calibration of the
Str\"omgren $\beta$ indices with spectroscopic effective temperatures
defined by the Fe I lines present in the optical spectra of slowly
rotating stars in the direction of Orion.
 The surface gravities for the rapid rotators were
assumed in Cunha et al. (1995) to be 4.0.  Here we have
adjusted the log g for HD~294297 to a higher value
(log g=4.4) because this produced an overall better fit in the
spectral region between 2494.5 and 2499.5 \AA. The sensitivity of the
derived boron abundance to changes in log g will be discussed in
Section 4.2.

Boron abundances were derived from spectrum synthesis of the region
around the B\,{\sc i} resonance line at 2496.771 \AA, which is the less
blended boron resonance line available.  In an effort to
place all derived boron abundances for near-solar metallicity stars on
the same absolute scale, the present analysis of the two Orion targets
is entirely consistent with the previous analysis of a sample of
near-solar metallicity solar type dwarfs (with [Fe/H] ranging from
-0.75 to +0.15) from HST archival data (Cunha et al.  2000).
The analysis of the B I spectral region in near solar-temperature
and solar-metallicity stars is complicated by the definition of the
continuum level; the spectra are crowded with strong absorption lines
and no regions are free from absorption. As discussed in previous
studies of B I (Cunha \& Smith 1999; Cunha et al. 1999; 2000), the
continuum level in this region was set for the Sun, where observed
specific intensities are available, using the combination of
synthesis code, line list, and solar model atmosphere. The relative
position of the continuum to spectral line depths does not vary
much over the limited range of T$_{eff}$ spanned by the Orion
G stars. The continuum level is allowed to vary slightly in order
to provide the best fit to the absorption lines.

This analysis technique was applied to 14 field F and G dwarfs
(Cunha et al. 2000),
for which reasonably accurate distances were available, and a
consistency check
was carried out between predicted and observed continuum fluxes. A
model surface flux at 2500 \AA, F$_{2500}$, comes from the model
atmosphere, and the predicted continuum flux observed
at the earth, f$_{2500}$, is
f$_{2500}$=F$_{2500}$(R/D)$^{2}$, where $R$ is the stellar radius and
$D$ is the distance; to compute the stellar radius requires distance,
apparent magnitude, reddening, a bolometric correction, and T$_{eff}$.
We call this the calculated continuum flux, f(calc). This flux
can be compared to the empirical continuum flux at 2500 \AA\ set by the spectrum
synthesis, which we call the observed continuum flux, f(obs).

The comparison of f(calc)/f(obs) for the field F and G dwarfs from
Cunha et al. (2000) was quite good (their Figure 3): with an
average f(calc)/f(obs) being 0.91$\pm$0.19 and having no trend in this
ratio with T$_{eff}$ over the range 5600-6700 K. This same technique
can be applied to the Orion stars as the distance to the Orion
association is reasonably well-known, e.g. Warren \& Hesser (1978);
de Zeeuw et al. (1999). The Orion association subgroups Ib and Ic
extend over a distance of $\sim$40 pc (de Zeuw et al. 1999) and we
adopt a single distance of 490 pc for the 3 G-dwarfs observed to
date by HST. A reddening of E(B-V)=0.05 is assumed for all Orion
stars and the (small) bolometric corrections are from B\"ohm-Vitense
(1989). Results for f(cal)/f(obs) in the Orion members are shown
versus T$_{eff}$ in the top panel of Figure 2, along with the previously derived
results for the field F and G dwarfs from Cunha et al. (2000). The
Orion members agree well with the field stars and indicate that
the analysis technique used to define the continuum is consistent.
The mean value of f(calc)/f(obs) for the Orion members is 0.89$\pm$0.18,
very similar to the field stars. The comparison between the
field-star sample and the Orion stars indicates that there are not
significant model-induced systematic differences between the two
sets of analyses. 

A quantitative estimate of the uncertainties associated with empirically
fitting the continuum levels can be made by investigating line depths 
as functions of T$_{eff}$ and [Fe/H] for the Orion targets, as well
as the larger sample of field dwarfs studied in Cunha et al. (2000).
Line depths measured relative to a continuum level should
depend primarily on the effective temperature and metallicity. 
The flux at a local high point in the spectra of these stars, near
2500 \AA\ (whose level is controlled by line blanketing), was then compared to 
the continuum flux derived empirically from
the spectrum synthesis (f(obs)) at this
wavelength. It is found that this
ratio of the observed line depth to the empirically derived continuum is a 
well-defined, smooth curve as a function of T$_{eff}$, for
near solar-metallicity stars (with [Fe/H]$\ge$-0.2).
This is shown in the
bottom panel of Figure 2 where we plot this ratio versus the effective
temperature for the field and Orion dwarfs. Inspection of the figure shows 
that this ratio varies from 
$\sim$0.25 at T$_{eff}$=5600K, to $\sim$0.65 at T$_{eff}=$6700K. Note
that the line-depth to continuum values increase for the metal-poor
stars, as expected due to reduced metal line blanketing. The observed 
scatter of the ratios about the mean curve defined for the near-solar
metallicity stars is $\pm$0.04. The flux ratio of
the local high point (at 2500 \AA) to the continuum varies from 0.3 to
0.4 for the three studied Orion stars and these three stars follow 
the relation of flux ratio versus T$_{eff}$ defined by the near-solar
metallicity field stars. 
If the continuum flux level for the Orion G-dwarfs is thus allowed to vary
by $\pm$0.04, the resulting derived B abundance varies by $\pm$0.11 dex:
this is a fair measurement of the abundance uncertainty introduced from
the uncertainties in the continuum level.

Our synthetic spectra were computed using the program LINFOR
(originally developed at Kiel University by H. Holweger, M. Steffen \&
W. Steenbock) and model atmospheres generated with the ATLAS9 code
(R. L. Kurucz, 1993 - private communication) for the stellar parameters in
Table 1.  The calculations of the synthetic spectra included a more
modern  value for the bound-free cross-section of Mg\,{\sc i}
3p$^{3}P^{o}$, which is the dominant source of continuous opacity in
the 2500 \AA\ region.  The importance of this choice is pointed out in
Cunha \& Smith (1999). In the calculation of the synthetic spectra the
adopted Mg abundance was scaled with the oxygen abundance for each star.
The adopted line list was compiled first in
order to fit the disk-center spectrum of the Sun, later this line list
was fine-tuned in order to fit the spectra of the sample of near-solar
metallicity dwarfs (mentioned above) with effective temperatures in
the range between 5650K and 6700K. This line list can be found in Cunha et al.
(2000).
In Figure 3 we illustrate synthetic spectra for the
studied stars.  Using a simple $\chi^{2}$ minimization we derive an
LTE boron abundance of log $\epsilon$(B)=2.60$\pm$0.20 for HD\,294297
and log $\epsilon$(B)=2.30$\pm$0.20 for BD\,-6$^{\circ}$\,1250 with
the uncertainties set by the sharpness of the $\chi^{2}$ minima.
Corrections for non-LTE effects in stars of solar temperatures and
metallicities are small: Kiselman \& Carlsson's (1996) non-LTE
calculations for the B\,{\sc i} lines indicate revised non-LTE abundances of
log $\epsilon$(B)=2.34 and log $\epsilon$(B)=2.65, respectively, for
HD\,294297 and BD\,-6$^{\circ}$\,1250.  The derived boron abundances for
the target stars plus their derived Li, Fe and O abundances are
assembled in Table 1.  We also added to this table other Orion members
that have been studied in previous papers (Cunha et al.
1995, 1998) and that will be brought into the discussion of the
abundance results and uncertainties that follow. We note the addition of
one star (P1374) that has not been published previously.

\bigskip
\section{Results and Discussion}

\subsection{The Abundances}

A summary of boron and oxygen abundances in Orion is presented in Figure
4. Combined with the
non-LTE B-star boron abundances\footnote{A large correction for non-LTE effects
on the B\,{\sc ii} 1362 \AA\ line
was  included (Cunha et al. 1997). The B\,{\sc iii} 2066 \AA\ line was
predicted to be minimally affected by departures from LTE. Proffitt et al.
(1999) observed and analyzed the 2066 \AA\ line in one of the four stars.
Our reanalysis (Lambert et al. 2000)
 of their spectrum with our model atmosphere gives a non-LTE
abundance in fair agreement with the non-LTE value
from the B\,{\sc ii} 1362 \AA\ line, bolstering our confidence in the
B\,{\sc ii} non-LTE corrections.}, the three points defined by the Orion
solar-type stars indicate clearly that there is no positive trend of
boron with oxygen in Orion. In fact, if there is any discernable trend
of B and O, it is an anticorrelation.  A possible anticorrelation is
made all the more convincing by the combination of two sets of results
from very different types of stars: B stars with T$_{\rm eff}$ $\sim$
18000-22000K (analyzing B\,{\sc ii}) and G stars with T$_{\rm eff}$ $\sim$
5850-6150K (analyzing B\,{\sc i}).  Taken together, the B and G stars seem to
define a single relation of decreasing boron with increasing oxygen
among the stellar members of Orion.
The most oxygen-rich and boron-poor Orion star in Figure 4 is
the G-star BD\,-5$^{\circ}$\,1317, whose B I spectrum was analyzed in Cunha
et al.
(1999). Much of the weight of a significant anticorrelation between B and
O falls on this star. Cunha et al. discussed whether the boron abundance
derived from B I could be influenced by some unkonwn effect, most
probably a chromosphere, but were unable to find an obvious explanation
for the weakened B I line in BD\,-5$^{\circ}$\,1317 (including the addition
of a
chromosphere to the model atmosphere), other than a low B abundance. This
star was studied for Li, Fe, and O in Cunha et al. (1995, 1998) and no
obvious spectral peculiarities were noted; however, future spectroscopic
studies of this star are encouraged.

If a linear least-squares fit is
performed on the log $\epsilon$(B) versus log $\epsilon$(O) data from
Orion, a slope of -1.1$\pm$0.2 is found (with a correlation coeficient r= -0.94),
indicating a significant decrease of B with O.

To probe the robustness of this apparent decrease of B as O increases, tests
on the boron and oxygen abundance dataset were conducted. As a first step,
the data point for BD\,-5$^{\circ}$\,1317 was excluded from another linear
least-squares fit on the remaining 6 points; in this case a significant
anticorrelation (r=-0.90) is still found between B and O with a slope of
-1.3$\pm$0.3 (within the errors, the same slope found with the inclusion of
BD\,-5$^{\circ}$\,1317). Reduced values of $\chi^{2}$ can also be computed
from the expression 
 
\begin{equation}
    \chi^{2}_{\mathrm{r}} = \frac{1}{\nu - 1} \Sigma \frac{(observed_{i} - fitted_{i})^{2}}{\sigma^{2}_{i}},
\end{equation}
where `observed' and `fitted' refer to the observed values and linear least-squares
computed values, respectively, $\sigma$ is the associated error, and $\nu$
is the number of degrees of freedom. Using $\sigma$=0.20 dex, the values of
$\chi^{2}_{\mathrm{r}}$ are 1.65 (with the inclusion of BD\,-5$^{\circ}$\,1317)
and 1.53 (with the exclusion of  BD\,-5$^{\circ}$\,1317): both values of
$\chi^{2}_{\mathrm{r}}$ indicate a good fit with a linear trend of 
log $\epsilon$(B) versus log $\epsilon$(O). If a zero slope between B and O
is assumed (i.e. no trend), the associated values of $\chi^{2}_{\mathrm{r}}$
are 15.40 (with the inclusion of BD\,-5$^{\circ}$\,1317) and 7.82 (with
the exclusion of  BD\,-5$^{\circ}$\,1317): both of these values indicate
extremely poor fits to the data (well past the 3$\sigma$ level of confidence).  
Because the observed trend is significant and opposite to
the global correlation between B and O in the Galaxy, over three
orders of magnitude in metallicity, an investigation into whether
errors can produce such an anticorrelation is carried out.

\subsection{Elemental Correlations and Stellar Parameter Errors}

As this paper deals with an analysis of G-type stars, the discussion
of possible errors in boron and oxygen abundances is confined to stars
of this spectral type for which the  B\,{\sc i} 2497 \AA\ line  and the
O\,{\sc i} 7770 \AA\ triplet provide the B and O abundances,
respectively.
  The ground-based and HST spectra analyzed for both oxygen and
boron are of relatively high S/N, and errors in the derived abundances
are due primarily to uncertainties in the crucial stellar parameters:
effective temperature (T$_{\rm eff}$), surface gravity (log g), and
microturbulent velocity ($\xi$).
Typical O\,{\sc i} equivalent widths for the three G stars were used as input
in order to compute the effects of the parameter changes on
derived O abundances.  For boron uncertainties, the discussions from
Primas et al.  (1999) and Boesgaard et al. (1998) were used.  The
results are that the oxygen abundances change by -0.08 dex for
$\Delta$T = +100K, +0.06 dex for $\Delta$log g = +0.3 dex, and -0.03
dex for $\Delta$$\xi$ = +0.2 km s$^{-1}$: the corresponding numbers
for boron are +0.10 (for $\Delta$T), -0.02 (for $\Delta$log g), and
-0.06 (for $\Delta$$\xi$) dex, respectively.
  Of special note are
the anticorrelated O and B abundance errors for changes in temperature
and surface gravity; for example, if the effective temperature of a
star were overestimated by 100K, the derived O abundance would be too
low by 0.08 dex, while the derived B abundance would be too high by
0.10 dex.  This effect results in spurious anticorrelations between B
and O for {\it random} temperature and gravity errors; however, we note
that the observed
scale of the anticorrelation over the range of derived B and O abundances
would require errors much larger than estimated,
for example, $\Delta$T $\simeq$ 600K.

Further constraints on the reality of a B-O anticorrelation are
provided by the Fe abundances from Cunha et al.  (1995; 1998) for 9
Orion members.  These stars span a T$_{\rm eff}$ range of 5600-6150K
and a tight distribution of Fe abundances is obtained with a spread
of $\pm$0.13 dex: this scatter is completely accounted for by the
observational errors, as argued by Cunha et al. Therefore, the Orion  F-G stars
show a single value of the Fe abundance, but not  a
single O abundance.
 For the Fe I lines used by Cunha et al.,
it is found that Fe abundances change by +0.08 dex for $\Delta$T=
+100K, -0.02 dex for $\Delta$log g=+0.3 dex, and -0.03 dex for
$\Delta$$\xi$= +0.2 km s$^{-1}$. Note that B I and Fe I have a similar
behavior. A 600K error in T$_{eff}$'s would clearly lead to a noticeable
spread in the
Fe abundances but that is not seen.

In order to test the magnitudes of spurious anticorrelations produced
by random T$_{\rm eff}$, log g, and $\xi$ errors, a program was used
to generate gaussian distributed noise of specified means and standard
deviations in these parameters, which were then used to compute errors
in boron, oxygen, and iron abundances.  Different starting model
abundances were used in order to understand under what conditions a
spurious anticorrelation, of the magnitude observed, between B and O
could be generated.  The fact that two very different types of stars
(spectral types B and G) seem to fall along a single relationship is
ignored here, and only uncertainties relevant to the G stars are
considered.
The internal abundance uncertainties from Cunha et al.  (1995; 1998)
and this study, for the Orion members, are dominated primarily by
uncertainties in the stellar parameters T$_{\rm eff}$, log g, and
$\xi$.  Estimates of these uncertainties are $\pm$150K in
T$_{\rm eff}$, $\pm$0.3 dex in log g, and $\pm$0.2 km s$^{-1}$ in
$\xi$. The estimated errors in stellar parameters are taken from
discussions in Cunha et al. (1995, 1998) for the Orion F/G dwarfs.
In these previous studies, based upon a comparison of photometric
and spectroscopic effective temperature scales, average differences
of $\pm$70K were found. We adopt a conservative approach here and double
these errors to $\Delta$T=150K. The uncertainties in surface gravity
and microturbulence are those from Cunha et al. (1995; 1998) and are not
as critical, as O I and B I are most sensitive to effective temperature.

A discussion of the abundance results begins with Fe and O in the
Orion G-stars from Cunha et al.  (1998) who found
that the observed Fe abundances could
be modelled as a single value, while the values of the derived O
abundances were too scattered to be explained by a single value.  This
contention is re-examined here using a different analysis.
 The error simulation technique
employed here consists of a beginning set of model data points that
represent the observed sample: in the case of Fe and O, this consists
of 9 points, while in the case of B and O it is 3 points.  An input
abundance distribution is assumed; for example, in the initial
modelling of Fe and O, a constant abundance value for each element in
the Orion members is tried.  Random T$_{\rm eff}$, log g, and $\xi$
errors are then generated for each input data point, resulting in
changes to the input Fe and O abundances ($\delta$Fe and $\delta$O)
which are then added to the input model abundances.  The input
abundances have now been perturbed by random stellar parameter errors.
Because this investigation is to probe possible correlations (or
anticorrelations) between pairs of elements, a linear least-squares
fit is then performed on the perturbed model points and a slope
derived: this slope can be compared to the corresponding slope derived
from the observed abundances.  The above procedure can be performed an
arbitrary number of times (each time is labelled as a
``realization''), with a slope computed for each realization.  A
distribution of slopes can then be constructed and compared to the
observationally derived slope.  The impact that stellar parameter
uncertainties can have on any underlying abundance correlations or
anticorrealtions can then be investigated.  In addition, standard
deviations in the abundances from the perturbed model points can be
compared to the observed values.  This exercise can thus test whether
the estimates of stellar parameter uncertainties are reasonable, as
well as the reality of possible correlations (or anticorrelations or
no correlations) between pairs of elements.

The top panel of Figure 5 shows the Fe versus O abundances as taken
from Cunha et al.  (1998).  No obvious trend exists and a linear
regression finds no correlation, with an insignificant positive slope
of 0.1$\pm$0.2 derived.  Note that in the previous discussion of the
various elemental abundance sensitivities to stellar parameters, the
Fe I and O I lines have both temperature and gravity sensitivties that
are of nearly equal magnitudes but in opposite senses, thus,
substantial, random stellar parameter errors would produce
anticorrelated Fe and O abundances.  This is clearly not observed and
suggests already that random errors are unlikely to be responsible for
the observed B-O anticorrelation.  However, an error simulation of the
Fe and O abundances can provide clues as to whether our error
estimates are reasonable, and how these errors translate into possible
effects on the B versus O trend.

The bottom panel of Figure 5 illustrates simulated data given
uncertainties in the stellar parameters: distributions of fitted
linear slopes to log $\epsilon$(Fe) versus log $\epsilon$(O) abundances
are shown (for 1000 realizations) using two different underlying model
abundances; a model with a single Fe and O abundance and a model with
single Fe and varying O. The slopes were derived from 9 input points
(as in the observed sample).  For the input model consisting of
constant Fe and O abundances we took the average log $\epsilon$(Fe)=
7.35 and log$\epsilon$(O)= 8.70, to which random abundance errors were
added given errors in T$_{\rm eff}$, log g, and $\xi$.  Because the Fe
I and O I lines are most sensitive to temperature, and in opposite
senses, a false anticorrelation is derived for the model points, which
is manifested in the bottom panel of Figure 4 as the distribution of
negative slopes centered on d(Fe)/d(O)= -0.8, reflecting the dominant
sensitivites to temperature of d(Fe)= +0.08 dex/100K and d(O)= -0.08
dex/100K. The average slope of -0.8, instead of -1.0 which would be
derived from temperature errors only, results from the lower
sensitivity of Fe I to gravity errors when compared to O I.

The observed slope of +0.1 (insignificant from zero slope) is far from
the derived average slope in the simulations of -0.80.  In addition,
using $\Delta$T$_{\rm eff}$ =150K, $\Delta$(log g)= 0.3 dex and
$\Delta$$\xi$= 0.2 km s$^{-1}$, the Fe scatter is found to be
$\pm$0.12 dex: in excellent agreement with the observed 0.13 dex.
Oxygen, on the other hand, is found to scatter in the simulations by
0.13 dex, far less than the observed 0.24 dex.  As found by Cunha et
al.  (1998), an intrinsic abundance scatter within the Orion members
of $\sim$0.5 dex is required.  If this oxygen spread is then included
in the input model data (while retaining a constant Fe abundance) the
second distribution of simulated slopes is shown in the bottom panel
of Figure 5 (again for 9 model points and 1000 realizations).  Here
the derived slopes from the simulated data are close to zero, as
observed in the real Orion members.  The difference between the slope
distributions in the two models is caused by the oxygen abundance
spread, as opposed to a single value.  With a spread in O, the
leverage to create a slope, produced by random errors, is lessened, as
well as the scatter in the derived slopes.  These simulations verify
the conclusions from Cunha et al.  (1998) that the Orion F- and
G-stars examined had a uniform Fe abundance, but significantly varying
O abundances (presumed to be due to selective enrichment of the Orion
Association from very massive SN II).

The trend of boron versus oxygen is now examined in light of the
results obtained for iron and oxygen.  Figure 6 illustrates the
simulations conducted on the B and O data, with the top panel again
showing the B versus O abundances derived for the Orion G-dwarf
members.  As was the case with Fe and O, if a single value for the O
and B abundances are assumed for all Orion members, random stellar
parameter errors will produce an apparent anticorrelation of B with O;
however, for realistic values of T$_{\rm eff}$, log g, and
microturbulence uncertainties, the model abundance scatter is much
smaller than that observed.  Inputting the O scatter of $\sim$0.5 dex,
as suggested by the Fe-O analysis, two model simulations are shown in
the bottom panel of Figure 6: each distribution of slopes is generated
from 3 model points (as in the observed sample), each run through 1000
realizations.  One input model had a constant B abundance, and the
distribution of slopes is scattered about a mean slope of 0.0, very
far from the observed slope of -1.  The second input model assumed an
anticorrelations of B with O, with log $\epsilon$(B) proportional to
log $\epsilon$(O)$^{-1}$.  In this case, with the given uncertainties
in stellar parameters, the slope distribution scatters about -1.0.
Note that with only 3 points, the scatter in slopes is larger than for
the 9 points used for the Fe and O model.  These simulations suggest
that the combination of Fe, O, and B abundances in the Orion members
indicate a constant Fe abundance, variable O abundance, and a variable
B abundance that is inversely proportional to oxygen.

Finally, lithium abundances for Orion members, when compared to oxygen
abundances, can further constrain the interpretation of boron in
Orion.  The top panel of Figure 7 shows the Li and O abundances for 10
Orion members.  Interpretation of Li must take into account the
susceptibility of this element to  destruction by warm protons.  Two separate
symbols are used in Figure 7: the filled symbols denote stars with lithium
near the
expected undepleted abundance of log $\epsilon$(Li)= 3.3, while the
open symbols are stars which have almost certainly depleted lithium  somewhat.
 In Cunha et al.  (1995), it was found that the
lower Li abundances in Orion members were for the slowest rotators (Vsin$\iota$
$\le$ 10 km s$^{-1}$) and may be stars in which lithium has been destroyed
from mixing
induced by rotational spindown. If the low Li stars are set aside,
 the very flat
values of undepleted Li versus O in the top panel of Figure 7 suggest
lithium is independent of oxygen.
Because the Li\,{\sc i}
line is very temperature sensitive (and in the opposite sense
to   O\,{\sc i} lines), the lack of a large scatter in the Li abundances
and the absence of an apparent
anticorrelation of Li and O, is an additional argument that large
errors in the stellar parameters do not exist for the Orion sample of
solar-type stars.  The bottom panel of Figure 7 shows distributions of
slopes in simulated data for two different models: constant Li with
varying O (resulting in model slope distributions near 0.0, as
observed in the real data), and Li abundances which decrease as
O$^{-1.0}$.  The observed Li abundances suggest no significant trend
of Li with O, and are most easily fit with the estimated errors of
150K in T$_{\rm eff}$, 0.3 dex in log g, and 0.2 km s$^{-1}$ in $\xi$.
With these errors, the derived anticorrelation of boron with oxygen is
real and must be explained.

Of interest to this investigation is a comparison
with the interstellar abundances of boron and oxygen
in the Association's gas. Both abundances have been measured
along sightlines in Orion but not to stars for which we have B and O
abundances. Meyer, Jura, \& Cardelli (1998) provided
the ``definitive'' interstellar O abundance including measurements for
5 stars in Orion. From these and other sightlines, the O
abundance measured from O\,{\sc i} 1356 \AA\ line was
remarkably uniform with no evidence for direct condensation of oxygen
atoms onto grains. When an estimate of oxygen atoms tied up in grains is
added to the measured abundance, the total oxygen abundance was
put at log $\epsilon$(O) = 8.60 with an uncertainty of less than 0.1 dex.

In diffuse interstellar clouds, gaseous boron is expected to be
predominantly present as B$^+$ ions.  Detections of these ions through
weak absorption at the 1362 \AA\ B\,{\sc ii} line are reported by Jura
et al.  (1996), Lambert et al.  (1998), and Howk, Sembach, \& Savage
(2000) for various sightlines.  Abundances, as summarized by Lambert
et al.  (1998), for sightlines to Orion are log $\epsilon$(B) $\simeq$
2.0 when independent observations of the H column density (mostly
H\,{\sc i}) are considered.  This is a lower limit to the interstellar
B abundance because some boron atoms may have condensed onto
interstellar grains.  The combination of measured interstellar
abundances in the direction of Orion of log $\epsilon$(B) $\simeq$ 2.0
and log $\epsilon$(O) = 8.6 (Meyer et al.  1998) is shown in Figure 4,
where it is seen to lie off the trend shown by the Orion stars.
Recently, Howk, Sembach, \& Savage (2000) have reported a higher
interstellar boron abundance, log $\epsilon$(B) $\simeq$ 2.4.  They
attribute their higher abundance to observations of clouds of lower
density in which depletion of boron onto grains is less severe.
However, the newly observed lines of sight are to distant stars,
primarily toward the Galactic center.  If the diffuse clouds are at a
similar distance, their boron abundance could differ from the local
(Orion) value on account of Galactic abundance gradients.  Therefore,
we adopt the local interstellar B abundance but recognize that it is a
lower limit.

\section{Supernovae and the Self-Enrichment of Molecular Clouds}

Assuming that the surface composition of a star reflects the
composition of the gas from which it formed, the observed variations
from star to star within an association can only be understood if the
parent cloud was chemically inhomogeneous and/or its composition
evolved and all the stars did not form at the same place and time.  To
interpret the various abundances of oxygen, iron and the light elements
in the Orion stars, one therefore has to determine the history and
geometry of the chemical enrichment, or if one prefers, the
distribution in both space and time of the different nucleosynthetic
episodes.

When a stellar association forms from the collapse of a chemically
roughly homogeneous cloud, the first-generation stars have
approximately the same composition.  The most massive stars evolve
quickly, on timescales of a few million years, and explode as SN II
which release in the ambient medium several solar masses of enriched
material, notably more than 1~\Msun~of pure oxygen per SN II. In
addition to this direct contamination of the gas, the SN II's have a
strong dynamical influence on the ambient medium: they produce a shock
wave which accelerates particles to relativistic energies and they
compress the surrounding gas.  Both theoretical models and direct
observation indicate that the explosion of a SN II within, or close
to, a molecular cloud can trigger the fragmentation of the gas and
lead to further star formation.  Depending on the mixing of the SN II
ejecta with the ambient, chemically unperturbed, gas the new stars
formed in the wake of previous SN II can show various O abundances,
bounded from below by the initial ISM O abundance and from above by
the O abundance in the ejecta.  The overabundance of oxygen in some
stars of an association can thus be attributed to the
\emph{self}-enrichment of the molecular cloud.

Note that in this model the O abundance varies in time, as more and
more SN II's explode and release O-rich material in the ambient medium,
but also from one place to another as the hazards of mixing and gas
fragmentation dictate.  Of course, one cannot expect to model the
Orion clouds in sufficient detail to determine the distribution of
abundances, density and other physical parameters over the whole
region, and we must limit ourselves to general trends and average
numbers.

To describe the variation of O abundances in Orion, we divide the gas
into two distinct components: the ejecta and the uncontaminated ISM,
which we simply call here the \emph{ambient medium}.  This simple model
considers addition of ejecta of mass $M_{\mathrm{ej}}$ to ambient
material of mass $M_{\mathrm{amb}}$ to provide a total mass $M_t =
M_{\mathrm{amb}} + M_{\mathrm{ej}}$ from which stars form.  If
$\alpha(X)$ denotes the mass fraction of element $X$ and $f =
M_{\mathrm{ej}}/(M_{\mathrm{ej}} + M_{\mathrm{amb}})$, the composition
of a star formed from the mixed gas of mass $M_t$ is given by

\begin{equation}
\alpha_*(X) = (1 - f)\alpha_{\mathrm{amb}}(X) + f\alpha_{\mathrm{ej}}(X).
\label{two}
\end{equation}

The relation between the B and O abundances in
Orion can be investigated using the simple model described above, in
which new stars form from various amounts of two chemically distinct
gas components (the SN II ejecta and the `ambient' medium) before they
are fully mixed and their compositions get homogenized.  Combining
Eq.~(2) for O and B, one can eliminate the mixing parameter, $f$, and
obtain:

\begin{equation}
    \alpha_{\star}(\mathrm{B}) = \alpha_{\mathrm{ej}}(\mathrm{B}) +
    K_{\mathrm{B/O}}[\alpha_{\star}(O) -
    \alpha_{\mathrm{ej}}(\mathrm{O})],
    \label{eq:B(O)}
\end{equation}
where
\begin{equation}
    K_{\mathrm{B/O}} = \frac{\alpha_{\mathrm{ej}}(\mathrm{B}) -
    \alpha_{\mathrm{amb}}(B)}{\alpha_{\mathrm{ej}}(\mathrm{O}) -
    \alpha_{\mathrm{amb}}(O)}.
    \label{eq:K-B/O}
\end{equation}

Alternatively, by exchanging $f$ and $1-f$ as well as
$\alpha_{\mathrm{ej}}(\mathrm{B})$ and
$\alpha_{\mathrm{amb}}(\mathrm{B})$ in Eqs.~(2)
and~(3), we have:

\begin{equation}
    \alpha_{\star}(\mathrm{B}) = \alpha_{\mathrm{amb}}(\mathrm{B}) +
    K_{\mathrm{B/O}}[\alpha_{\star}(O) -
    \alpha_{\mathrm{amb}}(\mathrm{O})],
    \label{eq:B(O)bis}
\end{equation}

From Eqs.~(3) or~(5), we see that the B
abundance in Orion stars can be either correlated or anticorrelated
with O, depending on whether the slope $K_{\mathrm{B/O}}$ is positive
or negative, respectively.  The anticorrelation reported in this
paper is thus compatible with our simple `mixing model' if
$\alpha_{\mathrm{ej}}(\mathrm{B}) < \alpha_{\mathrm{amb}}(\mathrm{B})$
(since $\alpha_{\mathrm{ej}}(\mathrm{O}) >
\alpha_{\mathrm{amb}}(\mathrm{O})$).  This is true if the
$\nu$-process for $^{11}$B production is negligible, since in that
case $\alpha_{\mathrm{ej}}(\mathrm{B})\rightarrow 0$.  Such a
conclusion would be very important for the light element
nucleosynthesis, since the $\nu$-process is generally invoked to
increase the B/Be and $^{11}$B/$^{10}$B ratios produced by standard
(high energy) nucleo-spallation processes (cf.
introduction).  If the Orion observations can be used to
rule out the $\nu$-process, then it seems inevitable that a low-energy
cosmic ray component (LECR) exists in the ISM, whose nature and origin
remains to be determined.

If we identify the most O-poor (log $\epsilon$(O) = 8.3) star
and the most O-rich (log $\epsilon$(O) = 9.1) star
as having formed from the (initial)
ambient and the most severely contaminated mixed gas, respectively,
$\alpha_*(\mathrm{O})/\alpha_{\mathrm{amb}}(\mathrm{O})$ is
given approximately by the ratio of the observed abundances of the most
oxygen-rich to oxygen poor stars ($\simeq 10^{0.8}
= 6.3$). (The qualification `approximately' is necessary because the
equation is framed in terms of mass fractions but the
spectroscopic analyses provide the  O/H  ratio subject to assumptions
about the chemical composition,
especially about
the He/H ratio, and, in the case
of the B\,{\sc i} line in F-G stars, about the Mg/H ratio (Cunha \& Smith
1999).)

In order for  B and O to be anticorrelated,
the SN\,II ejecta must have a B/O ratio that is
less than the ambient material. The limiting case obviously occurs when
the ejecta are thoroughly depleted in B and rich in O, a condition
that denies the $\nu$-process a significant role in the synthesis of B.
In such a case,

\begin{eqnarray}
\alpha_*(B) = (1 - f)\alpha_{\mathrm{amb}}(B).
\label{two}
\end{eqnarray}

The maximum observed value of $\alpha_*(\mathrm{O})/\alpha_{\mathrm{amb}}(\mathrm{O})$
$\sim$ 6.3 can be used to set a lower limit on the ratio
$\alpha_{\mathrm{ej}}(\mathrm{O})/\alpha_{\mathrm{amb}}(\mathrm{O})$
(the lower limit would be this value if the star were composed purely 
of SN II ejecta).
If a ratio $\alpha_{\mathrm{ej}}(\mathrm{O})/\alpha_{\mathrm{amb}}(\mathrm{O})$ is assumed,
Eq.~{2} gives
an estimate of $f$ that may be used in Eq.~{6} to calculate
the reduction in the B abundance between the initial ambient gas and the
most heavily contaminated star. Obviously, the greater the ratio
$\alpha_{\mathrm{ej}}(\mathrm{O})/\alpha_{\mathrm{amb}}(\mathrm{O})$, the smaller
the reduction of the B abundance.
For an arbitrary value of $\alpha_{\mathrm{ej}}(\mathrm{O})/\alpha_{\mathrm{amb}}(\mathrm{O})$ = 7 
(slightly larger than the lower limit
of 6.3) and $\alpha_*(\mathrm{O})/\alpha_{\mathrm{amb}}(\mathrm{O}) = 6$ (see above), $f$ =
0.83 and the observed B abundance in such a star would be 
$\alpha_*(B) = 0.17\alpha_{\mathrm{amb}}(B)$
(a reduction of about 0.8 dex, which is roughly that observed). A value of 
$\alpha_{\mathrm{ej}}(\mathrm{O})/\alpha_{\mathrm{amb}}(\mathrm{O}) \simeq 7$ is
not an unreasonable value for a Type II supernova, where 1$M_\odot$ of
oxygen may be synthesized and ejecta may amount to 10$M_\odot$. In this
model, there is no constraint that the B-reduction must approximately
equal the O-increase. For example, if 
$\alpha_{\mathrm{ej}}(\mathrm{O})/\alpha_{\mathrm{amb}}(\mathrm{O})$ is raised to 10, the B reduction
is only 0.25 dex.
This exercise does not address the feasibility of
retention of an adequate mass of SN\,II ejecta by gas that will
subsequently form the stars with oxygen abundances above ambient values.
This interesting issue which was aired by Cunha \& Lambert (1992) may
be bypassed here; a more compelling challenge to this simple explanation of
the O versus B anticorrelation must be faced.

In the simple mixing model presented here, if B is not synthesized in significant 
amounts  by the $\nu$-process, then possibly $^{7}$Li is not produced either.
In such a case, the Li and B abundances should be
correlated and decline in tandem with increasing O abundance.
Lithium is not detectable in B-type stars. Lithium and oxygen
abundances of the F-G stars were derived by Cunha et al. (1995,
1998). Figure 7 shows that the lithium abundances are nearly independent
of the
O abundances, with perhaps a slight decrease. A simple, but by no means
unique,
interpretation  is that synthesis of Li accompanies the O synthesis.
However, the interpretation of
Li abundances is complicated by the possibility that Li is destroyed within
stars. It is important to note that in Figure 7 no undepleted Li abundances
are measured
below log $\epsilon$(O)=8.7. There is a tendency for the Orion stars
with lower oxygen abundances to have lower Li abundances: these stars
are older, tend to rotate more
slowly, and have probably depleted their initial Li abundances (Cunha et al.
1995).
In order to properly place
Li within the context of the evolution of B and O in Orion, it will
be necessary to measure the behavior of undepleted Li abundances in Orion
members with lower oxygen abundances.

Although the absence of a significant $\nu$-process is a sufficient
condition for a local anticorrelation between B and O, we draw
attention to the fact that it is not a necessary condition.  In a
model advocated by Parizot (2000), energetic particles (SBEPs)
accelerated inside a superbubble created by repeated SN\,II's in an OB
association may be responsible for synthesis of Li, Be, and B in a
supershell.  The geometry of the Orion-Eridanus superbubble (see e.g.
Burrows et al.  1993) is such that the Orion clouds themselves can
actually be considered as part of the supershell.  Indeed, the history
of star formation in Orion indicates that the molecular cloud is
`eroded' from the side which faces the Orion-Eridanus superbubble, and
a star formation wave propagates deeper and deeper into the cloud. 
Now this star-forming side of the cloud is irradiated by SBEPs
(Parizot 1998) and an overabundance of spallation products like Be and
B is to be expected there.  Therefore, the `ambient gas' component in
Eq.~(2) could be very much enriched in Li, Be and B. This makes it
possible that the resulting B abundance, $\alpha_{\mathrm{amb}}(B)$,
is greater than $\alpha_{\mathrm{ej}}(B)$ even in the presence of a
significant $\nu$-process.  From the astrophysical point of view, it
depends on the penetration of the SBEPs inside the Orion clouds.  The
more they penetrate, the less $\alpha_{\mathrm{amb}}(B)$, since the B
production is then distributed over a larger volume.

\section{Perspective}

As recalled in the introduction, theoretical studies of light element
production have up to now confined themselves to the interpretation of
the \emph{general} increase of LiBeB abundances as a function of
metallicity.  On local scales, complex behaviors may result from the
fact that CNO and LiBeB are not necessarily produced at exactly the
same place or at the same time.  Indeed, before the local production
of B and O, say, effectively results in a global increase of the B and
O abundances in the ISM, the O-rich and B-rich material have to mix
together and with the rest of the ISM. Now if new stars form before,
or during this mixing episode, one should expect them to show quite
unusual compositions.  Therefore, it is not obvious that the
\emph{local} B-O anticorrelation found in Orion is contradictory with
the results obtained on the Galactic scale.  In a recent study,
Parizot \& Drury (2000) have addressed the question of a possible
scatter in the Be/O and B/O ratios in the Galaxy, across a mean value
accounted for by the superbubble model.  The idea was very similar:
because O and BeB are not produced (or released) together in
superbubbles, various elemental ratios can actually result from
inhomogeneous mixing of the O-rich gas with the BeB-rich material.
Observational evidence of such a scatter in the Be data is available
in Boesgaard et al.  (1999).

Detailed calculations of the expected values for
$\alpha_{\mathrm{amb}}(B)$ and $\alpha_{\mathrm{ej}}(B)$, as well as
the implications of the Orion data for the SB model will be presented
in Paper~II. Here, we have noted that the observed B-O anticorrelation
can be accounted for within a simple self-enrichment model.  First, if
the $\nu$-process is negligible, the anti-correlation is an inevitable
prediction of the model.  Second, if the $\nu$-process proves to be
significant, the observed anticorrelation simply means that
$\alpha_{\mathrm{amb}}(B) > \alpha_{\mathrm{ej}}(B)$, which is at
least plausible considering the very high local production of B from
SBEPs (see Paper~II for more details and quantitative estimates).
Finally, it is important to realize that the Orion observations can
actually be used to constrain the light element production models.
Although the B-O anticorrelation alone cannot be used to determine the
weight of the $\nu$-process in the Galaxy, a joint study of the Be
abundances should give decisive information.  Since Be is not expected
to be a $\nu$-process product and is then solely a fruit of
spallation, it can be used to set the parameters of the SB and the
mixing models.  Any difference between Be and B can then be attributed
to the $\nu$-process, and serve to quantify it.
We predict that Be will also be found anticorrelated
with O in the Orion association, and \emph{even more} than B. This is
because B is produced by two means: nucleo-spallation, which leads to
the anticorrelation, and neutrino-spallation, which alone would lead
to a B-O positive correlation, since the boron produced in this way is
`ready-mixed' with the oxygen in the SN II ejecta.  Since the
$\nu$-process is irrelevant for Be, the `slope' of the Be-O
anticorrelation must be at least equal, and possibly greater (in
absolute value) than that of the B-O anticorrelation.  Note that the
slopes are here the ones described by Eq.~(3), namely
$|K_{\mathrm{B/O}}|$ and $|K_{\mathrm{Be/O}}|$ (which should actually
be normalized by dividing them by $(\mathrm{B/O})_{\odot}$, say, and
$(\mathrm{Be/O})_{\odot}$, respectively).  This is different from the
slope seen in Figure~4, where the abundances are plotted on
logarithmic scales.  It is remarkable, however, that the only
anticorrelation in logarithmic variables which is compatible with a
anticorrelation in linear variables (as predicted by the mixing model,
Eq.~(3), is one of slope $-1$, which is what we observe for Orion.

Finally, the case of lithium is more complicated, since the Galactic
evolution models indicate that the spallation processes may not be the
main contributors, and AGB stars probably produced most of the Li
currently observed in the Galaxy.  The absence of a clear correlation
or anticorrelation found in Figure 7 could thus result from the fact
that the $\nu$-process is roughly balanced by the SBEP-induced
production, or by the fact that none of these processes modifies
significantly the Li abundance in the ejecta and the ambient medium.

\bigskip
We thank Ivo Busko for help in issues related to the STIS spectra.
This research is supported in part by NASA through the
contract NAG5-1616, and the grant
GO-06520.01.95A from the Space Telescope Science Institute, which is
operated by the Association of Universities for Research in Astronomy,
Inc., under NASA contract NAS5-26555.  We also acknowledge support
from the National Science Foundation through grant AST96-18459.
EP was supported by the TMR programme of the European Union under
contract FMRX-CT98-0168.

\clearpage

\figcaption[Cunha.fig1.ps]{The final STIS spectra of BD-6$^{\circ}$\,1250 (top
panel) and HD294297 (bottom panel). The brightness is in units of
ergs cm$^{-2}$ s$^{-1}$ \AA$^{-1}$ arcsec$^{-2}$. The solid angle
(arcsec$^{2}$) subtended by one pixel corresponds to 8.41x10$^{-4}$.
\label{fig1}}

\figcaption[Cunha.fig2.ps]{{\bf Top panel}: Ratio of the predicted continuum
flux to the observed continuum 
flux (at $\lambda$=2500 \AA) for field F and G dwarfs from Cunha et al.
(2000) and the Orion G-stars. Predicted fluxes are derived using
the synthesis code LINFOR with ATLAS9 model atmospheres, plus estimated
stellar radii, distances, reddenings, luminosities, and T$_{eff}$'s.
{\bf Bottom panel}: The ratio of the observed line depth to the empirically
derived continuum as a function
of the effective temperatures. 
The filled circles represent the Orion targets while the open circles and x's
are the FG dwarfs studied by Cunha et al. (2000). The more metal
poor stars (represented by x's) have line depths which are closer to
the continuum. The continous curve is simply a third-order polynomial fit to 
the near-solar metallicity field dwarfs.
\label{fig2}}

\figcaption[Cunha.fig3.ps]{The observed (filled squares) and synthetic
spectra for BD-6$^{\circ}$\,1250 (top panel) and
HD294297 (bottom panel). The best fit boron abundances of respectively 2.3
and 2.6 are shown as solid curves. The synthetic spectra for both stars
were calculated for a microturbulence velocity of 1.3 kms$^{-1}$.
\label{fig3}}

\figcaption[Cunha.fig4.ps]{The behavior of boron versus oxygen in Orion.
The four B-type stars are represented by filled squares and the three
studied solar-type stars are shown as filled circles. The oxygen abundances
for the B-stars are from Cunha \& Lambert (1994) and for the G-stars
are from Cunha et al. (1998). Also shown 
are the interstellar medium value observed in the direction of Orion
(Lambert et al. 1998) and the solar value. There is an obvious anticorrelation
of boron and oxygen in Orion.
\label{fig4}}

\figcaption[Cunha.fig5.ps]{The top panel shows the abundances of iron
versus oxygen in 9 Orion-member G-dwarfs from Cunha et al. (1998).  The
points with filled circles are the 3 members for which boron abundances
have been determined from HST spectra of B I.  The bottom panel summarizes
the error simulations for Fe and O by showing linear slope distributions
derived from input model abundances of Fe and O.  Two model distributions
are shown, each generated from 9 input points to which random T$_{\rm eff}$,
log g, and $\xi$ errors are used to calculate the corresponding errors in
log $\epsilon$(Fe) and log $\epsilon$(O).  Linear least-squares fits were
then carried out on the resultant Fe and O abundances.  The slope
distributions consist of 1000 realizations for each input model.  The model
with slopes centered on negative values of -0.8 consisted of input
abundances which had constant values of Fe and O abundance for each
input point: this distribution of slopes is a poor representation of the
observed slope of $\Delta$Fe/$\Delta$O= +0.1$\pm$0.2.  In addition, a
single O-abundance for the Orion members leads to a $\sigma$(O)= $\pm$0.13
dex, only half that observed in the real stars.  The second model was one
in which Fe was constant, while O ranged in abundance over 0.5 dex; this
model is represented by the slope distribution centered near an Fe/O slope
of 0.0.  This is in good agreement with the observed value, as well as
reproducing the standard deviation of the observed O abundances.
\label{fig5}}

\figcaption[Cunha.fig6.ps]{This figure is similar to Figure 4; the top panel
shows the boron versus oxygen abundance for the 3 Orion G-dwarfs observed by
HST.  The bottom panel summarizes the error simulations using 3 input
points, each run 1000 times with random temperature, gravity, and
microturbulent errors to produce perturbed B and O abundances.  Linear
least-square fits were then used to generate slopes, which are shown in
the bottom panel as distributions.  The slope distribution which is centered
near $\Delta$B/$\Delta$O$\sim$ -0.25 is generated from an input, underlying
abundance distribution of constant boron, with an oxygen abundance spread
of 0.5 dex: this model is not able to reproduce the observed slope of $\sim$
-1 between B and O.  The second slope distribution near a slope of -1 is
derived from an input model abundance set in which B is proportional to
O$^{-1}$ and this model fits well the observed slope, as well as the
observed standard deviations of both the B and O abundances.
\label{fig6}}

\figcaption[Cunha.fig7.ps]{This figure is similar to both Figures 4 and 5.
The lithium versus oxygen abundances for 10 Orion members listed in
Table 1 are shown in the
top panel. The oxygen abundances are from Cunha et al. (1998) and the Li
abundances are from Cunha et al. (1995). The open circles represent stars 
in which mild Li depletion
has occurred from the undepleted Orion abundance of log $\epsilon$(Li)=
3.2.  The bottom panel shows the results for error simulations of Li
and O consisting of slope distributions from Li versus O derived
from 10 input points, each run 1000 times with random T$_{\rm eff}$,
log g, and $\xi$ errors, for two input abundance distributions.  One
model has a constant Li abundance and an O spread of 0.5 dex (the distribution
with slopes centered near -0.15), while the other assumes that Li declines
with O$^{-1}$ (as found for boron).  The observed slope (-0.17$\pm$0.20)
of Li versus O is derived only from the 6 stars with undepleted Li
and this slope is matched by the model with constant Li with O.
\label{fig7}}

\newpage
.
\vspace{5 cm}
\center TABLE 1
\center Stellar Parameters and Abundances
\vspace{10 pt}
\begin{small}
\begin{tabular}{ccccccccc}
\hline
\hline
Star   & T$_{eff}$(K)  & Log g & Log $\epsilon$(Li)$_{nlte}$   &
Log $\epsilon$(B)$_{lte}$   &
Log $\epsilon$(B)$_{nlte}$  &
Log $\epsilon$(O)$_{nlte}$  &
Log $\epsilon$(Fe)$_{lte}$ \\
\hline
P1179     & 6050 & 4.0 & 3.13 & --  & --  & 8.71 & --   \\
P1374    & 6390 & 4.0 & 3.24 & --  & --  & 9.00 & 7.46 \\
P1455     & 5950 & 4.0 & 3.26 & --  & --  & 8.86 & 7.59 \\
P1657     & 6100 & 3.8 & 2.56 & --  & --  & 8.72 & 7.20 \\
P1789     & 6120 & 4.0 & 3.06 & --  & --  & 9.29 & 7.31 \\
P1955     & 5890 & 4.0 & 3.18 & --  & --  & 9.13 & 7.53 \\
P2374     & 5900 & 3.9 & 2.30 & --  & --  & 8.78 & 7.41 \\
BD-5$^{\circ}$\,1317 & 5850 & 4.0 & 3.11 & 2.0 & 2.06 & 9.11 & 7.34 \\
BD-6$^{\circ}$\,1250 & 5950 & 4.0 & 2.74 & 2.3 & 2.34  & 8.74 & 7.33 \\
HD294297  & 6150 & 4.4 & 2.56 & 2.6 & 2.65 & 8.61 & 7.32 \\
\hline
\end{tabular}
\end{small}
\newpage

\begin{figure}
\centerline{\epsfig{file=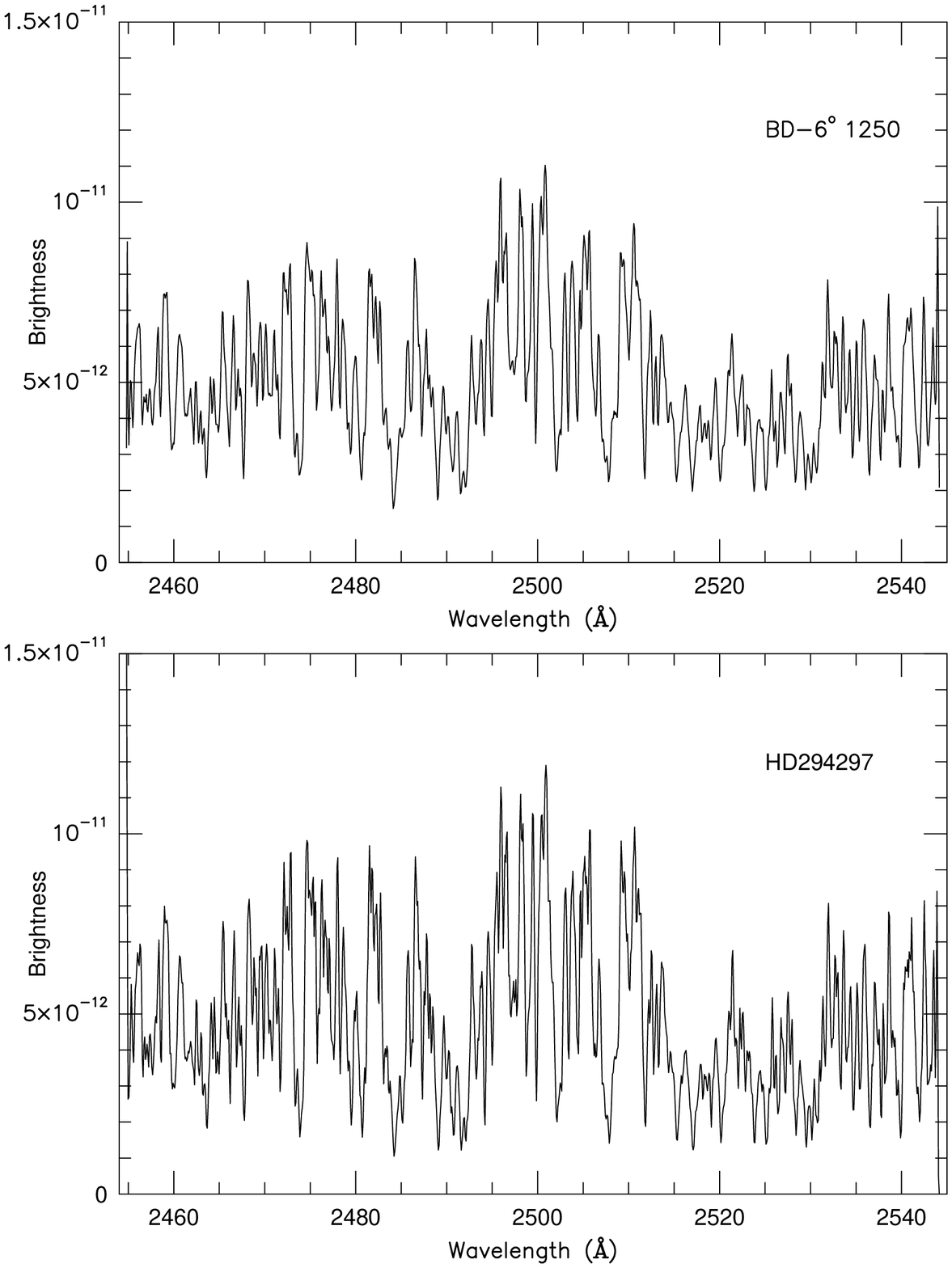,width=\textwidth,clip=}}
\end{figure}

\begin{figure}
\centerline{\epsfig{file=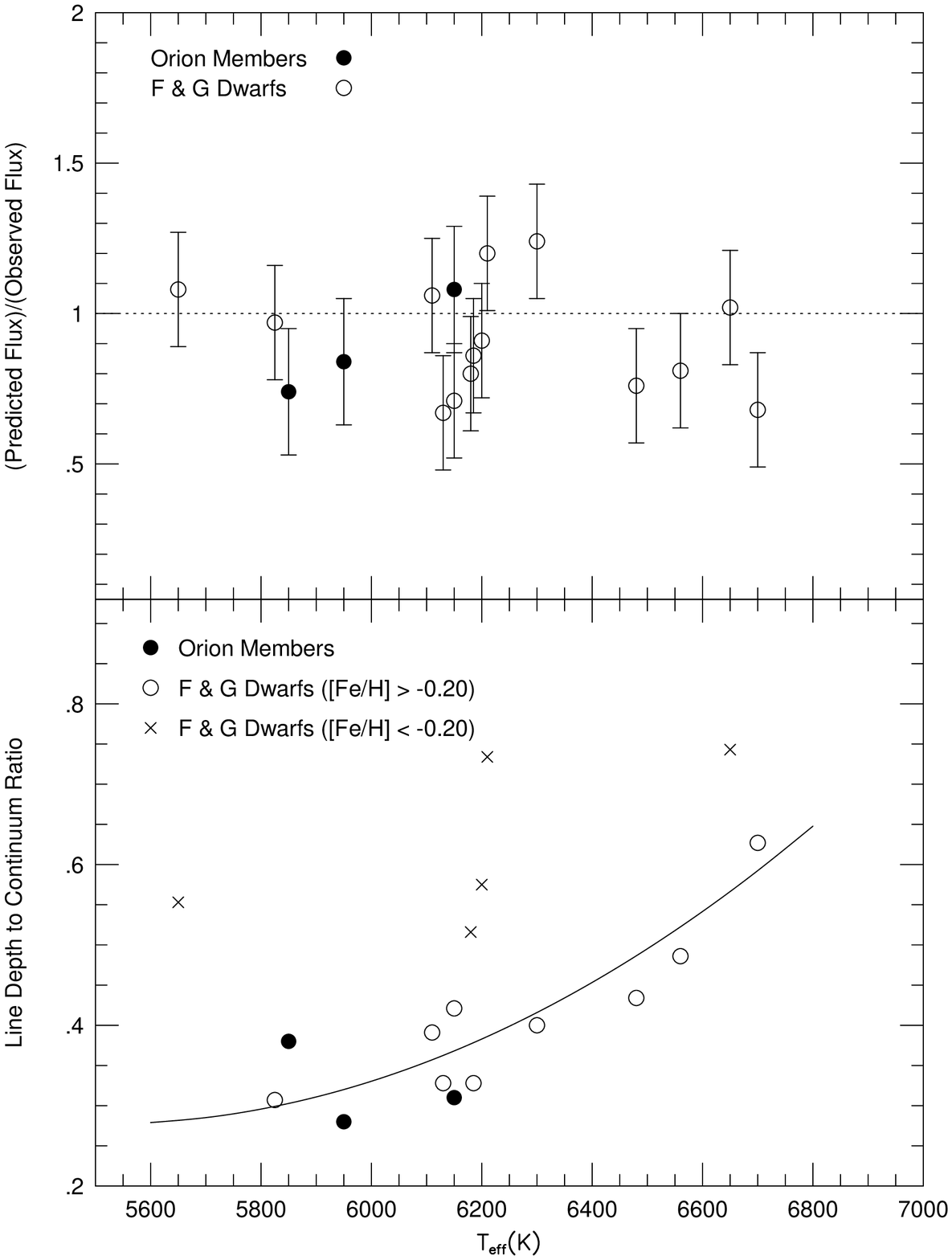,width=\textwidth,clip=}}
\end{figure}

\begin{figure}
\centerline{\epsfig{file=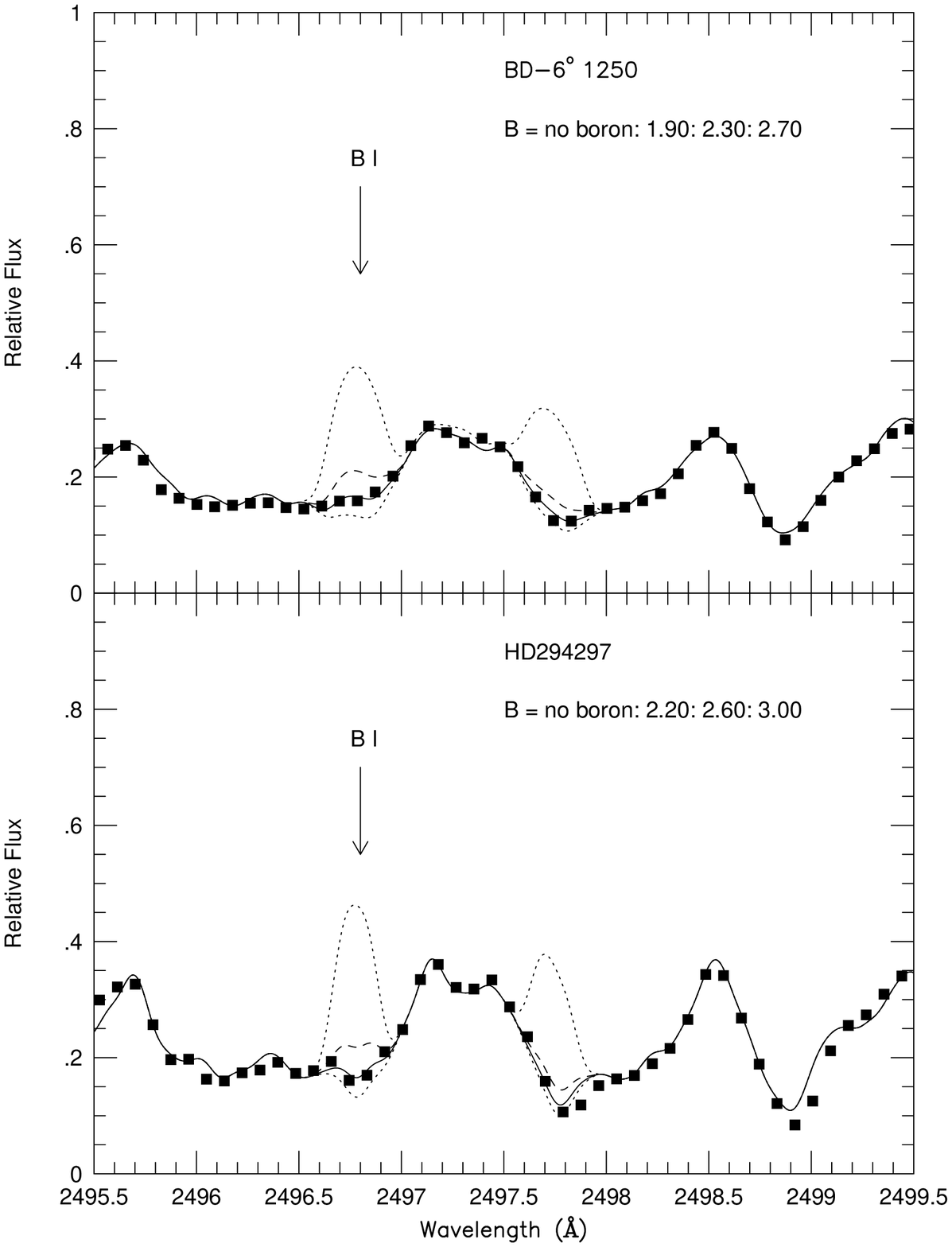,width=\textwidth,clip=}}
\end{figure}

\begin{figure}
\centerline{\epsfig{file=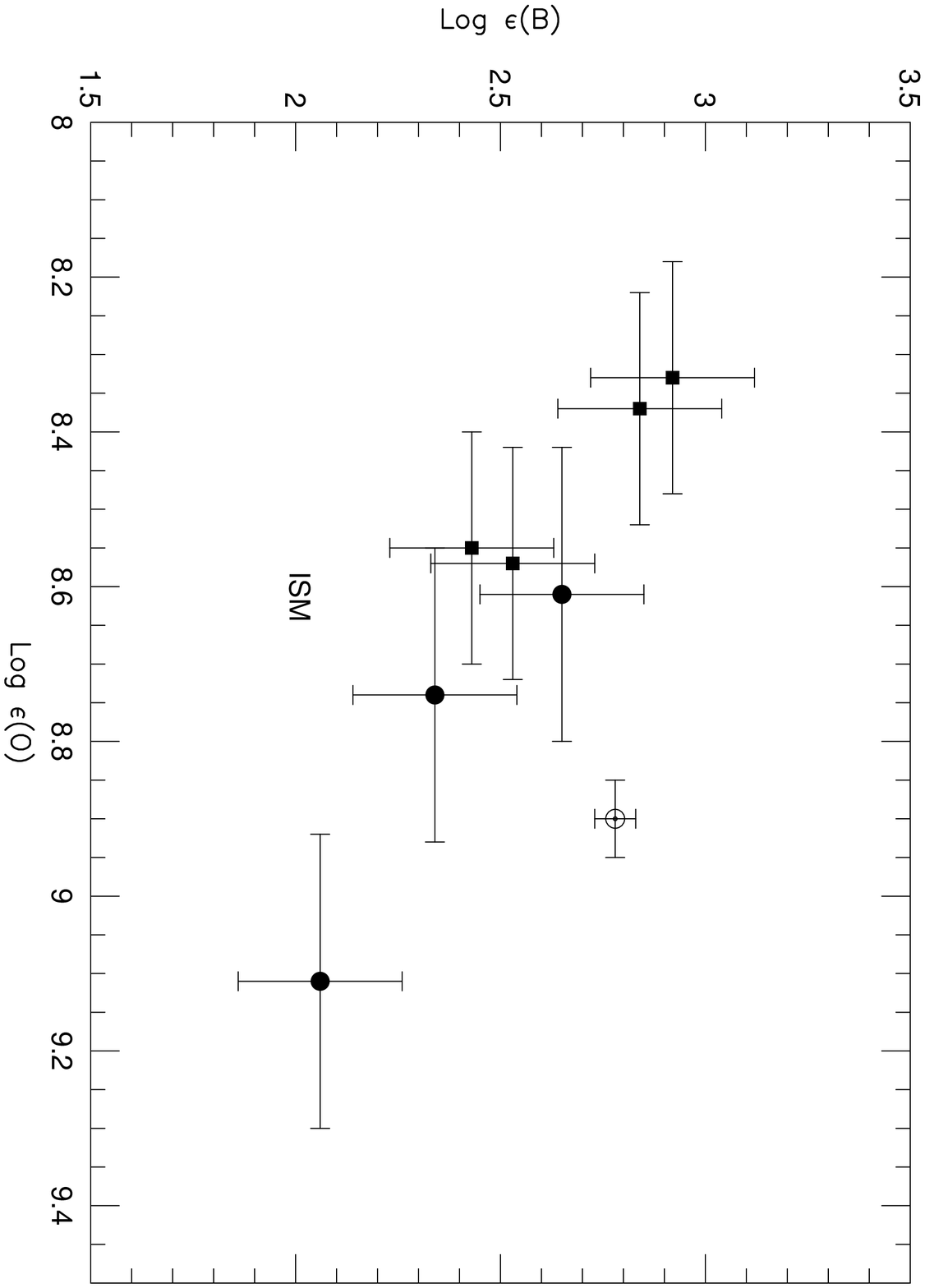,width=\textwidth,clip=}}
\end{figure}

\begin{figure}
\centerline{\epsfig{file=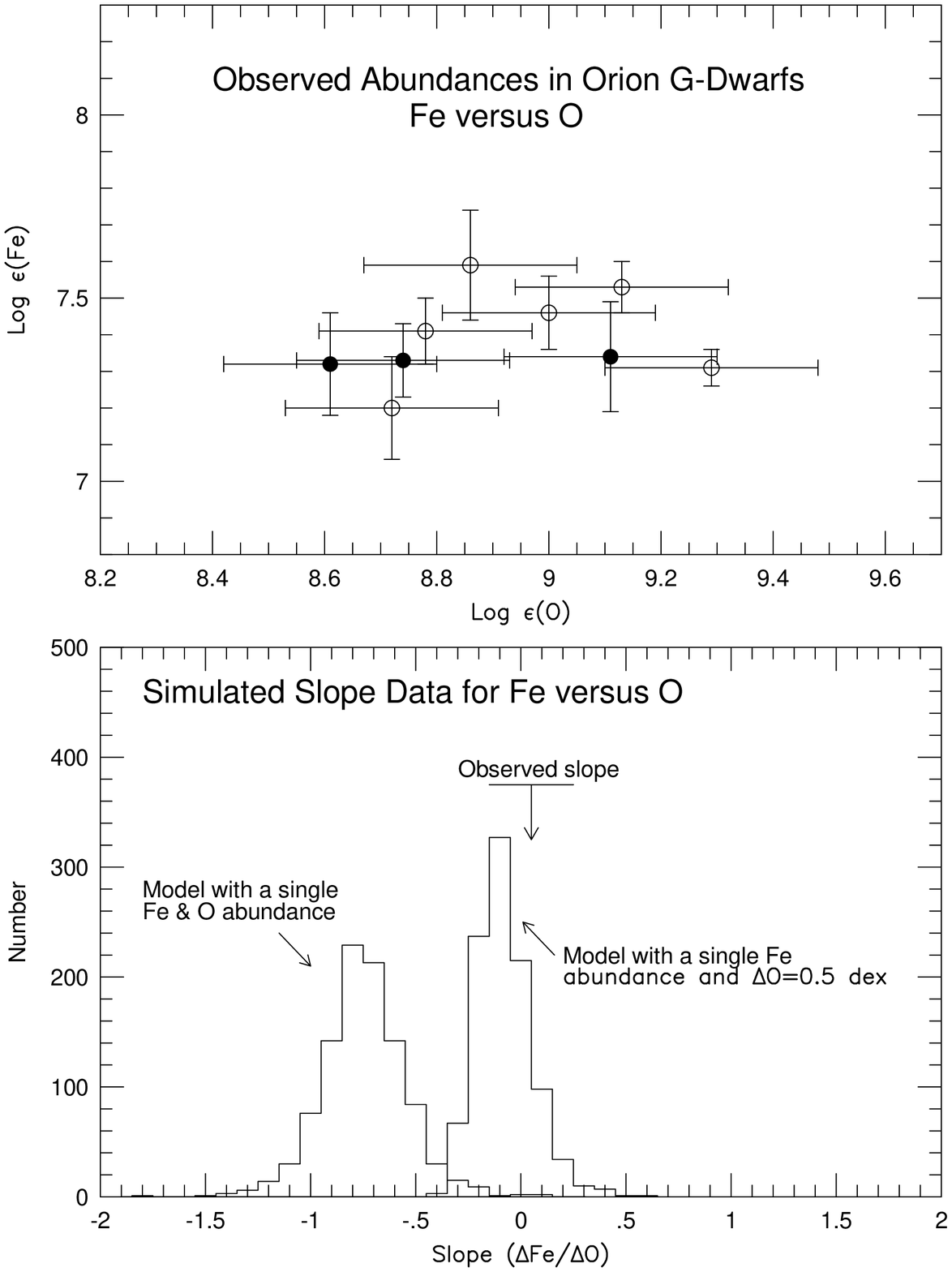,width=\textwidth,clip=}}
\end{figure}

\begin{figure}
\centerline{\epsfig{file=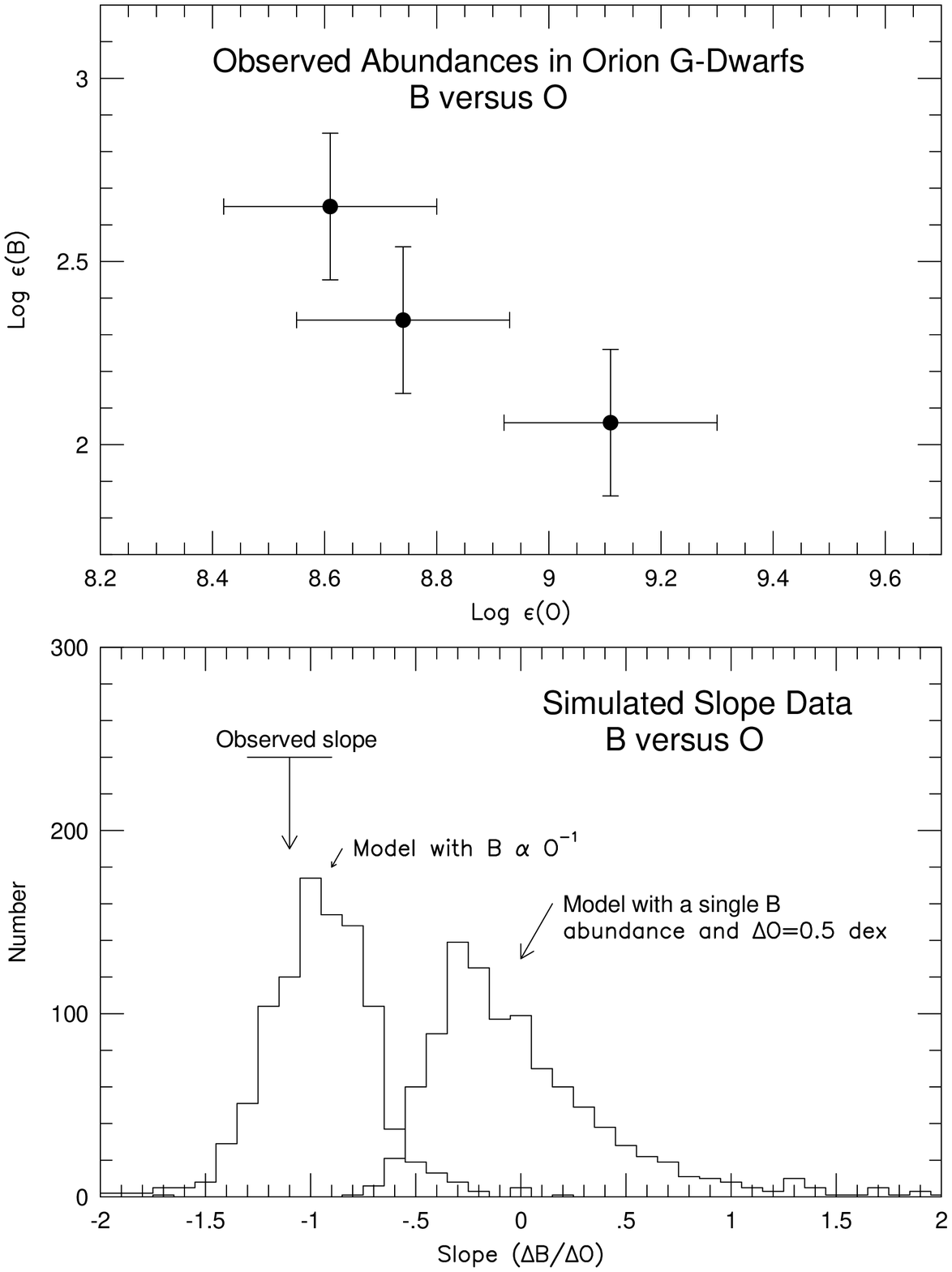,width=\textwidth,clip=}}
\end{figure}

\begin{figure}
\centerline{\epsfig{file=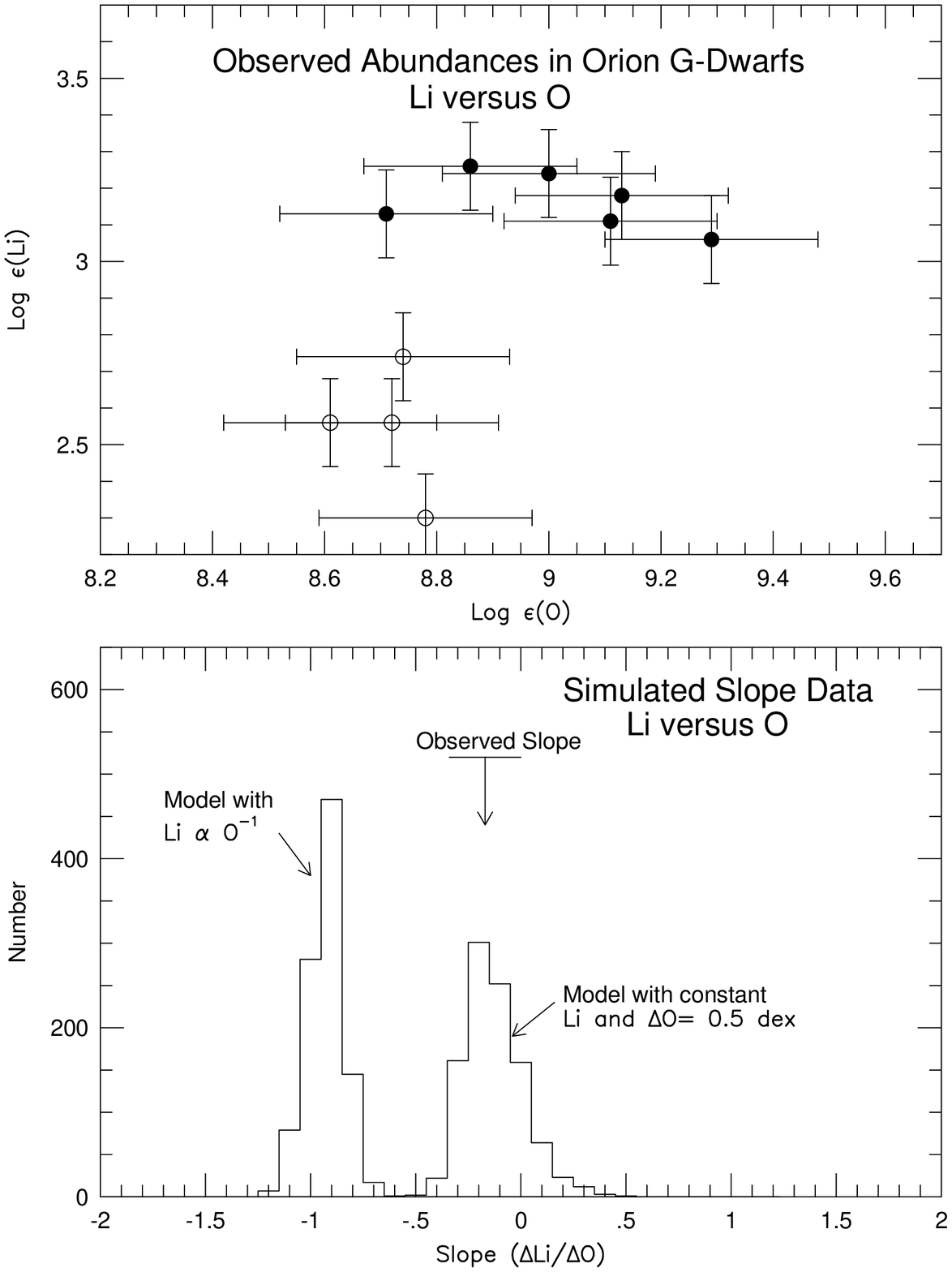,width=\textwidth,clip=}}
\end{figure}

\end{document}